\documentclass[letterpaper, 10 pt, conference]{ieeeconf}  

\IEEEoverridecommandlockouts                              

\usepackage{amsmath} 
\usepackage{amssymb}  
\usepackage{caption}
\usepackage{subcaption}
\usepackage{graphicx}
\usepackage{algorithm}
\usepackage{algpseudocode}
\usepackage[utf8]{inputenc}
\usepackage{hyperref}
\usepackage{subcaption}
\usepackage{multirow}
\usepackage{multicol}
\usepackage{cite} 
\usepackage{tikz}
\usepackage{tikz-3dplot}
\usepackage{subcaption}
\usepackage{tabularx}
\usepackage{booktabs}    
\usepackage{array}       

\title{\LARGE \bf
Analysis of Optimal Thrust to Mass Ratio Requirement for Maximizing Payload Mass of Lunar Landing Mission
}

\author{Aditya Rallapalli$^{1}$, Suraj Kumar$^{1}$, Rijesh MP$^{1}$, Dr. C K Koteswar Rao$^{1}$, Bharat Kumar GVP$^{1}$
\thanks{$^{1}$The authors are associated with U R Rao Satellite Center, Indian Space Research Organization, Bengaluru, Karnataka, India\{adityar, surajk, rijesh, ckrao, bharat\}@ursc.gov.in}
}

\begin{document}

\maketitle
\thispagestyle{empty}
\pagestyle{empty}

\begin{abstract}

Recent successful lunar landing missions have generated significant interest among space agencies in establishing a permanent human settlement on the Moon. Building a lunar base requires multiple and frequent landing missions to support logistics and mobility applications. In these missions, maximizing payload mass—defined as the useful cargo for human settlement—is crucial. The landing mass depends on several factors, with the most critical being the maximum thrust available for braking and the engine's specific impulse (ISP). Generally, increasing engine thrust for braking reduces flight duration and, consequently, gravity losses. However, higher thrust also introduces trade-offs, such as increased engine weight and lower ISP, which can negatively impact payload capacity. Therefore, optimizing the descent trajectory requires careful consideration of these parameters to achieve a global solution that maximizes payload mass.Most existing research focuses on solving optimal control problems that minimize propellant consumption for a given thrust. These problems are typically addressed through trajectory optimization, where a minimum-fuel solution is obtained. The optimized trajectory is then executed onboard using polynomial guidance.
In this paper, we propose an outer-layer optimization approach based on a Pareto-optimal solution. This method iterates on the maximum available thrust for descent trajectory optimization while incorporating a penalty/loss function that accounts for engine mass and ISP losses. By applying this approach, we identify a globally optimal solution that maximizes payload mass while ensuring an optimal landing trajectory.

\end{abstract}

\section{Introduction}
Lunar exploration has garnered significant interest among various space agencies, with the goal of establishing a permanent human presence on the Moon and ensuring long-term survivability. The Artemis III mission aims to send astronauts by 2028, marking the first crewed lunar landing since Apollo 17. China’s Chang’e mission plans to send astronauts by 2030, while India’s first successful lunar landing \cite{rallapalli2024landing} has paved the way for future exploration, with a target to send its first astronaut by 2040. These missions will focus on long-term sustainability and the establishment of a permanent lunar settlement.

A sustained human presence on the Moon requires frequent heavy-duty cargo deliveries to support astronauts. Therefore, maximizing payload mass is a key objective. In this context, payload mass is defined as the cargo a lander can transport to the lunar surface, while any system that does not contribute to cargo delivery is considered dead weight.

Extensive research has been conducted on optimal trajectory design\cite{accikmecse2013lossless}, primarily focusing on minimizing propellant consumption for a given thrust-to-mass ratio. Trajectory optimization seeks to determine the optimal powered descent start point relative to the landing site, minimizing propellant usage for a given thrust. These milestones are tracked using closed-loop onboard guidance, which generates a suboptimal flight performance \cite{rijeshdesign}\cite{chakrabarti2024convex}.

For lunar landing missions, high-thrust deep-throttling engines \cite{cheatham1966apollo} are generally desirable to achieve fuel-optimal trajectories. However, higher thrust comes with trade-offs, such as increased engine mass and lower specific impulse (ISP), which ultimately reduce the payload capacity. These losses must be accounted for within an optimization framework to determine a globally optimal payload-maximizing solution, known as a Pareto-optimal solution.

In this paper, we propose an outer-layer optimization framework that extends traditional trajectory optimization by incorporating engine thrust and ISP losses. This approach enables the generation of a global solution aimed at maximizing payload mass.
The paper is organized as follows: Section II discusses the dynamics of a typical lunar landing problem and trajectory design considerations. Section III presents the optimization framework for fuel-optimal trajectory design, incorporating engine mass as a constraint. Section IV analyzes simulation results and findings with conclusion and future scope in section V.
\begin{figure}[htbp]
\centering
\begin{subfigure}{0.48\columnwidth}
\centering
\tdplotsetmaincoords{70}{120}
\begin{tikzpicture}[scale=1.2,tdplot_main_coords]
\draw[thick] (0,0,0) circle (1);
\draw[->] (0,0,0) -- (1.5,0,0) node[below]{$X$};
\draw[->] (0,0,0) -- (0,1.5,0) node[right]{$Y$};
\draw[->] (0,0,0) -- (0,0,1.5) node[above]{$Z$};
\tdplotsetcoord{P}{10}{60}{40}
\draw[thick,->] (0,0,0) -- (P) node[midway, above=3pt]{$r$};
\fill[red] (P) circle (1.5pt);

\coordinate (Rp) at (Pxy); 
\coordinate (Pxy) at ({1.2*cos(40)*cos(60)},{1.2*cos(40)*sin(60)},0);
\draw[dashed] (0,0,0) -- (Pxy);
\tdplotdrawarc[->]{(0,0,0)}{0.5}{0}{60}{anchor=north}{$\theta$} 
\foreach \t in {0,5,...,40} {
    \pgfmathsetmacro{\X}{0.9*cos(\t)*cos(60)}
    \pgfmathsetmacro{\Y}{0.9*cos(\t)*sin(60)}
    \pgfmathsetmacro{\Z}{0.9*sin(\t)}
    \coordinate (A\t) at (\X,\Y,\Z);
}
\draw[->,black] (A0) -- (A5) -- (A10) -- (A15) -- (A20) -- (A25) -- (A30) -- (A35) -- (A40)
node[right] {$\phi$};
\draw[->,dashed] (P) -- ++(-1,-1.,0) node[below]{$x$};
\draw[->,dashed] (P) -- ++(-0.4,0.4,0) node[right]{$y$};
\draw[->,dashed] (P) -- ++(1,1,1) node[above]{$z$};
\end{tikzpicture}
\caption{Spacecraft orbiting the moon}
\end{subfigure}
\begin{subfigure}{0.48\columnwidth}
\centering
\tdplotsetmaincoords{70}{120}
\begin{tikzpicture}[scale=1.2,tdplot_main_coords]
\draw[->] (0,0,0) -- (1.5,0,0) node[below]{$x,u$};
\draw[->] (0,0,0) -- (0,1.5,0) node[right]{$y,v$};
\draw[->] (0,0,0) -- (0,0,1.5) node[above]{$z,w$};
\tdplotsetcoord{T}{4}{55}{40}
\draw[thick,->] (0,0,0) -- (T) node[above right] {$T$};
\coordinate (Tp) at ({2*cos(40)*cos(50)},{2*cos(40)*sin(50)},0);
\draw[dashed] (0,0,0) -- (Tp); 
\tdplotdrawarc[->]{(0,0,0)}{0.4}{0}{50}{anchor=north}{$\beta$}
\tdplotsetthetaplanecoords{40} 
\tdplotdrawarc[tdplot_rotated_coords,->,thick]
  {(0,0,0)}{0.6}{0}{55}{pos=0.1,above right}{$\alpha$}

\end{tikzpicture}
\caption{Thrust vector in body frame}
\end{subfigure}
\caption{Spherical Coordinate Frame for dynamics}
\label{fig:coord_frame}
\end{figure}
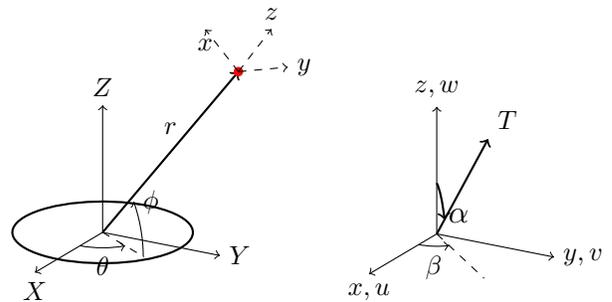
\section{Lunar Lander Translational Dynamics}
Lunar lander translational dynamics can be expressed in moon centered moon fixed reference frame using spherical co-ordinate reference. The translational dynamics for trajectory optimization is considered in the moon-centered inertial frame. The 3D equations of motion are given by,
\begin{align}
\label{traj_opt_dyn}
    \dot{r} &= w \nonumber \\
    \dot{\theta} &= \frac{u}{rcos\phi} \nonumber \\
    \dot{\phi} &= \frac{v}{r} \nonumber \\
    \dot{w}  &= \frac{Tsin\beta}{m} - \frac{\mu}{r^2}  + \frac{(u^2 + v^2)}{r} \\ 
    &+ (-2u\omega cos\phi + r\omega^2 cos^2\phi) \nonumber \\
    \dot{u} &= \frac{Tcos\alpha cos\beta}{m} + \frac{(-uw + uv tan\phi)}{r} \\
    &+ (-2w\omega cos\phi + 2v\omega sin\phi) \nonumber \\
    \dot{v} &= \frac{(Tsin\alpha cos\beta)}{m} \frac{(-vw - u^2 tan\phi)}{r} \\
    &+ (-2u\omega sin\phi - r\omega^2 sin\phi cos\phi) \nonumber \\
    \dot{m} &= -\frac{T}{I_{sp}g_0}
\end{align}

Here, $r$ represents the radial distance from the moon centre, $\theta, \phi$ represents the lander latitude and longitude respectively; $\omega$ represents moon rotational velocity; $u,v,w$ represents the tangential, across and radial velocity components respectively; $m$ represents the mass of the lander; $\mu$ represents the moon's gravitational constant; $I_{sp}$ represents specific impulse of the engine and $g_0 = 9.81$. The thrust vector is parameterized in the body frame by its magnitude $T$, declination angle, $\beta$, and right ascension angle $\alpha$. 

Generally, global trajectory is segmented into a multi-phase trajectory composed of rough braking phase, fine braking phase, and terminal descent phase \cite{rallapalli2024landing,kumar2025powered}. These milestones are derived based on mission and sensor constraints. Here we will focus on a single phase trajectory design for inner loop optimization.

\section{Fuel optimal trajectory design}

The objective of trajectory optimization is the minimization of propellant consumption which,
equivalently, is posed as the maximization of vehicle mass at the final time as,
\begin{equation}
J = -m(t_f)
\end{equation}
The state variables are given as $[r, \theta, \phi, w, u, v, m]^T$ and the control variables are $[T, \alpha, \beta]^T$. The dynamics governing the motion are given in Equation~\eqref{traj_opt_dyn}.

The initial conditions at \( t_0 \) are:
\begin{align}
    r(t_0) &= r_0, \quad \theta(t_0) = free, \quad \phi(t_0) = free  \\
    w(t_0) &= w_0, \quad u(t_0) = u_0, \quad v(t_0) = v_0  \\
    m(t_0) &= m_0
\end{align}

The final conditions at \( t_f \) are:
\begin{align}
    r(t_f) &= r_f, \quad \theta(t_f) = \theta_f, \quad \phi(t_f) = \phi_f  \\
    w(t_f) &= 0, \quad u(t_f) = 0, \quad v(t_f) = 0
\end{align}
where \( (r_f, \theta_f, \phi_f) \) corresponds to the target landing site.

The control constraints are:
\begin{align}
    0 \leq T \leq T_{\max}  \\
    |\alpha| \leq \alpha_{\max}, \quad |\beta| \leq \beta_{\max}
\end{align}

It can be seen that in following trajectory optimization framework maximum thrust available is considered as fixed. Also, the cost function does not include effective payload mass. 

\section{Pareto optimal framework}

Classical trajectory optimization problem doesn't address the problem of finding optimal thrust required to maximize the payload mass. Here, we propose the following augmentation to the trajectory optimization problem by modifying the equations of motion and cost function to explicitly consider the mass, thrust and ISp engines. 

\begin{itemize}
    \item Case 1:
    Consider a case where maximum thrust from engine is fixed but number of engines can be varied to generate higher thrust required for trajectory optimization. Then updated cost function is given as,
\begin{equation}
J = -(m(t_f) - \sum_{i=1}^{n} m_{eng,i}) 
\end{equation}
where $m_{eng,i}$ is the dry mass of $i^{th}$ engine and $n$ is the total number of engines. 
The updated mass flow rate dynamics is given as,
\begin{equation}
    \dot{m} = -\sum_{i=1}^{n}\frac{T_i}{I_{sp,i}g_0}
\end{equation}

 where $T_i, I_{sp,i}$ is the thrust and ISp from $i^{th}$ engine. The maximum thrust, $T_{max}$ is given as 
\begin{equation}
    T_{max} = \sum_{i=1}^{n} T_{max,i}
\end{equation}
where $T_{max,i}$ is the maximum thrust from $i^{th}$ engine. \\

The number of engine,$n$, is the additional variable in the optimization. 
\end{itemize}

\begin{itemize}
    \item Case 2:
    Consider a single engine is used and maximum thrust from engine can be increased with a quadratic penalty on mass and ISp based on maximum thrust engine can deliver
\begin{equation}
J = -(m(t_f) - m_{eng}) 
\end{equation}
where $m_{eng}$ is the dry mass of the engine which can be defined as
\begin{equation}
    m_{eng} = m_{0} + C_{1}T_{max} + C_{2}T_{max}^2   
\end{equation}
Similarly ISp of the engine can be defined as 

\begin{equation}
    I_{sp} = I_{sp0} + D_{1}T_{max} + D_{2}T_{max}^2   
\end{equation}
 the co-efficient $C_{1}$, $C_{2}$, $D_{1}$ and $D_{2}$ can  be derived based on engine characteristics/ experimental data.

 Following problem can now be solved using non-linear programming.

\end{itemize}
 
\section{Simulation Results}

\begin{table}[t]
\centering
\caption{Simulation Parameters}
\begin{tabular}{|l|c|}
\hline
\textbf{Parameter} & \textbf{Value} \\
\hline
Lunar gravity (\(g\)) & \(1.62 \, \text{m/s}^2\) \\
Initial mass (\(m_0\)) & \(4000 \, \text{kg}\) \\
Initial altitude & \(30000 \, \text{m}\) \\
Initial vertical velocity & \(0.0 \, \text{m/s}\) \\
Initial horizontal velocity & \(1688 \, \text{m/s}\) \\
Final altitude & \(800 \, \text{m}\) \\
Final velocity & \(0 \, \text{m/s}\) \\
\hline
\end{tabular}
\label{tab:sim_params}
\end{table}

\begin{table}[t]
\centering
\caption{Case 1 Engine Parameters}
\begin{tabular}{|l|c|}
\hline
\textbf{Parameter} & \textbf{Value} \\
\hline
Engine Max Thrust ($T_{max}$) & \(900 \, \text{N}\) \\
Engine ISp       ($I_{sp}$) & \(310 \, \text{s}\) \\
Engine Mass & \(8 \, \text{kg}\) \\
\hline
\end{tabular}
\label{tab:sim_params}
\end{table}

\begin{table}[t]
\centering
\caption{Case 2 Engine Parameters}
\begin{tabular}{|l|c|}
\hline
\textbf{Parameter} & \textbf{Value} \\
\hline
Engine Mass Coefficient  &  \\
    \hspace{1cm} $M_{0}$ & 2.229 \\
    \hspace{1cm} $C_{1}$ & 0.006288 \\
    \hspace{1cm} $C_{2}$ & -7.109e-08 \\
Engine Isp Coefficient  ($I_{sp}$) &  \\
    \hspace{1cm} $I_{sp0}$ & 311.3 \\
    \hspace{1cm} $D_{1}$ & -0.0005976 \\
    \hspace{1cm} $D_{2}$ & 4.755e-09 \\
\hline
\end{tabular}
\label{tab:sim_params}
\end{table}

\begin{figure}[t]
  \centering
  \includegraphics[width=0.5\textwidth]{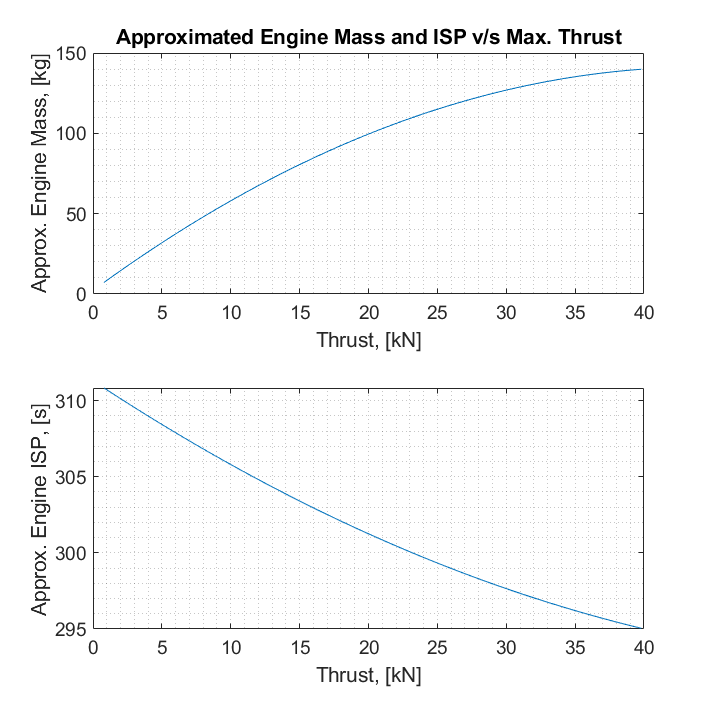}
  \caption{Mass and ISp of engine as function of maximum throttling}
  \label{fig:eng_mass_isp}
\end{figure}

The simulation was carried out using a set of parameters representative of a typical lunar landing scenario. Table 1 represents the list of parameters assumed for trajectory design and arriving at optimal thrust requirement. Table 2 shows the typical engine parameters taken for study to arrive at optimum number of engines which maximizes payload mass. Table 3 shows the engine characteristics as defined in equation [17] and [18], used to arrive at optimum thrust for maximizing payload mass. Figure [1] shows the engine mass and ISp as a function of maximum thrust engine can deliver which is again used for pareto studies.

\begin{figure}[h]
  \centering
  \includegraphics[width=0.5\textwidth]{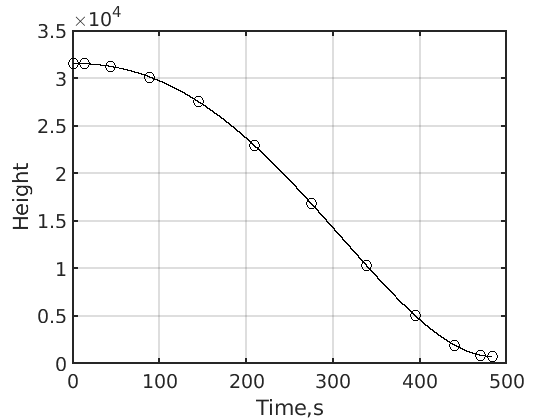}
  \caption{Altitude Profile}
  \label{fig:ht}
\end{figure}

\begin{figure}[h]
  \centering
  \includegraphics[width=0.5\textwidth]{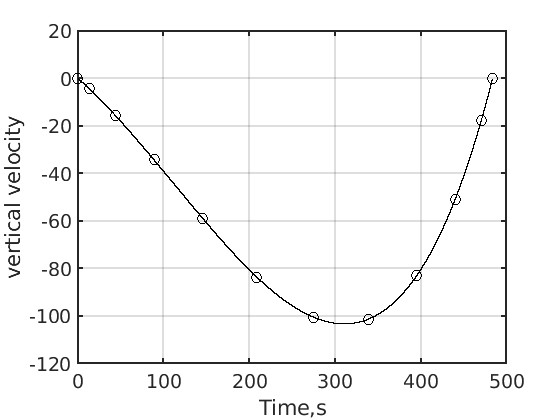}
  \caption{Vertical Velocity Profile}
  \label{fig:vv}
\end{figure}

\begin{figure}[h]
  \centering
  \includegraphics[width=0.5\textwidth]{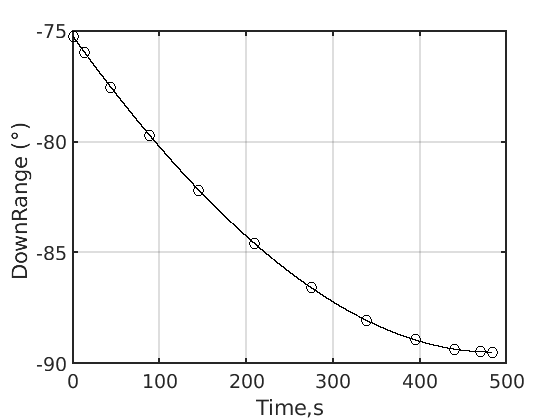}
  \caption{Downrange profile}
  \label{fig:dr}
\end{figure}

\begin{figure}[h]
  \centering
  \includegraphics[width=0.5\textwidth]{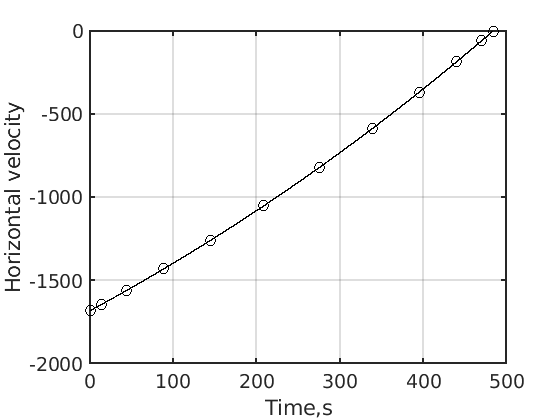}
  \caption{Horizontal Velocity profile}
  \label{fig:hv}
\end{figure}

\begin{figure}[h]
  \centering
  \includegraphics[width=0.5\textwidth]{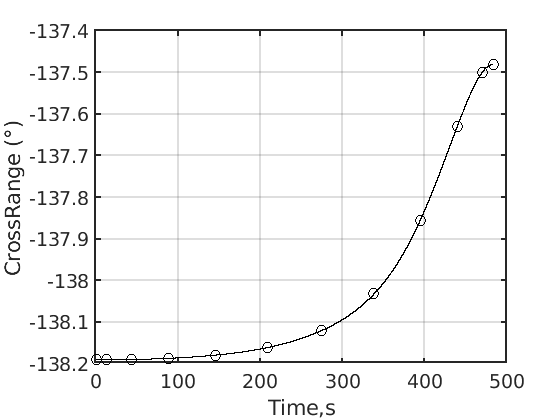}
  \caption{Crossrange profile}
  \label{fig:cr}
\end{figure}

\begin{figure}[h]
  \centering
  \includegraphics[width=0.5\textwidth]{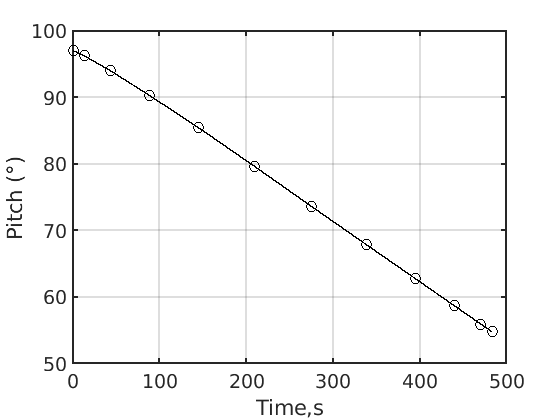}
  \caption{Pitch profile}
  \label{fig:pitch}
\end{figure}

\begin{figure}[h]
  \centering
  \includegraphics[width=0.5\textwidth]{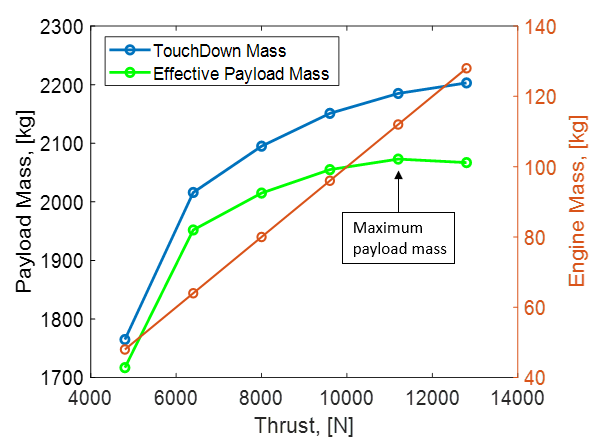}
  \caption{Pareto optimal curve}
  \label{fig:poc}
\end{figure}

Figures \ref{fig:ht} through \ref{fig:pitch} present the typical trajectory parameters—altitude, vertical velocity, downrange distance, horizontal velocity, and pitch angle—as functions of time, obtained through trajectory optimization. Both cases were optimized using nonlinear programming, and Figure \ref{fig:poc} summarize the overall results.

It can be observed that the payload mass, which was maximized using classical trajectory optimization, exhibits an increasing trend with higher maximum thrust, eventually approaching saturation. However, the effective payload mass, as defined in Equations [13] and [17], initially increases with maximum thrust but then begins to decline beyond a certain point. This clearly indicates the existence of an optimal thrust value that maximizes the effective payload mass.

The results suggest that if the lunar lander is designed with the thrust level identified through Pareto analysis, the effective payload mass can be maximized. It is interesting to note that optimal thrust mass ratio obtain from the study is ~3 $m/s^2$ which is mass independent. We observe the literature of thrust to mass ratio used in various landing mission is close to the number observed through this study \cite{graves1972apollo} \cite{li2016guidance}.

\section{Conclusions}

In this paper, we have extended the classical trajectory optimization framework by incorporating additional system-level parameters, specifically engine thrust, specific impulse (Isp), and engine mass. This enhancement enables a more comprehensive analysis aimed at determining the optimal engine thrust that maximizes the effective payload mass delivered to the lunar surface.

Our studies show that the effective payload mass exhibits a global maximum when the maximum thrust available during powered descent is systematically varied. This behavior highlights the presence of an optimal thrust level beyond which further increases in thrust result in diminishing returns or even a reduction in effective payload capacity due to increased engine mass and associated penalties.

By integrating these parameters into the trajectory optimization process, the proposed framework allows for the identification of the thrust configuration that yields the most favorable trade-off between engine performance and payload delivery. Consequently, this approach provides valuable insight for the preliminary design and sizing of lunar landers, ensuring that the effective payload mass is maximized within the constraints of propulsion system characteristics. 
\pagebreak
\section*{Acknowledgments}
The authors would like to express their sincere gratitude to UR Rao Satellite Centre for providing the necessary support and resources throughout the course of this work. We are especially thankful to the Director, Shri Sankaran M, Deputy Director Shri Sudhakar S, Group Director Dr. M S Siva and Dr. Ravikumar L their continued encouragement and guidance. We also extend our heartfelt appreciation to the domain experts and technical teams whose valuable insights and constructive feedback significantly contributed to the refinement and success of this study.


\end{document}